\documentclass[preprint2,times]{aastex61}
\usepackage{graphicx}

\shorttitle{Damping of sunspot umbral waves}
\shortauthors{Krishna Prasad et al.}

\begin{document}
\title{The Frequency-dependent Damping of Slow Magnetoacoustic \\ Waves in a Sunspot Umbral Atmosphere}

\correspondingauthor{S. Krishna Prasad}
\email{krishna.prasad@qub.ac.uk}

\author[0000-0002-0735-4501]{S. Krishna Prasad} 
\affiliation{Astrophysics Research Centre, School of Mathematics and Physics, Queen's University Belfast, Belfast, BT7 1NN, UK}                                  
\author{D. B. Jess} 
\affiliation{Astrophysics Research Centre, School of Mathematics and Physics, Queen's University Belfast, Belfast, BT7 1NN, UK}                                  
\affiliation{Department of Physics and Astronomy, California State University Northridge, Northridge, CA 91330, U.S.A.}                                  
\author{T. Van Doorsselaere}
\affiliation{Centre for mathematical Plasma Astrophysics, Mathematics Department, KU Leuven, Celestijnenlaan 200B bus 2400, B-3001 Leuven, Belgium}                                  
\author{G. Verth}
\affiliation{School of Mathematics and Statistics, The University of Sheffield, Hicks Building, Hounsfield Road, Sheffield, S3 7RH, UK}
\author{R. J. Morton}
\affiliation{Department of Mathematics, Physics and Electrical Engineering, Northumbria University, Ellison Building, Newcastle upon Tyne, NE1 8ST, UK}
\author{V. Fedun}
\affiliation{Department of Automatic Control and Systems Engineering, University of Sheffield, Sheffield, S1 3JD, UK}
\author{R. Erd\'elyi}
\affiliation{Solar Physics \& Space Plasma Research Centre (SP2RC), School of Mathematics and Statistics, University of Sheffield, Sheffield S3 7RH, UK}
\affiliation{Department of Astronomy, E\"otv\"os Lor\'and University, Budapest, P.O.Box 32, H-1518, Hungary}
\author{D. J. Christian}
\affiliation{Department of Physics and Astronomy, California State University Northridge, Northridge, CA 91330, U.S.A.}

\begin{abstract}
High spatial and temporal resolution images of a sunspot, obtained simultaneously in multiple optical and UV wavelengths, are employed to study the propagation and damping characteristics of slow magnetoacoustic waves up to transition region heights. Power spectra are generated from intensity oscillations in sunspot umbra, across multiple atmospheric heights, for frequencies up to a few hundred mHz. It is observed that the power spectra display a power-law dependence over the entire frequency range, with a significant enhancement around 5.5{\,}mHz found for the chromospheric channels. The phase-difference spectra reveal a cutoff frequency near 3{\,}mHz, up to which the oscillations are evanescent, while those with higher frequencies propagate upwards. The power-law index appears to increase with atmospheric height. Also, shorter damping lengths are observed for oscillations with higher frequencies suggesting frequency-dependent damping. Using the relative amplitudes of the 5.5{\,}mHz (3 minute) oscillations, we estimate the energy flux at different heights, which seems to decay gradually from the photosphere, in agreement with recent numerical simulations. Furthermore, a comparison of power spectra across the umbral radius highlights an enhancement of high-frequency waves near the umbral center, which does not seem to be related to magnetic field inclination angle effects.
\end{abstract}


\section{Introduction}
Slow magnetoacoustic waves (SMAWs) are in general compressive in nature, which makes them easily detectable through imaging observations. With the advent of high-resolution observations from ground- and space-based instruments, both standing and propagating SMAWs have been discovered in a multitude of structures in the solar atmosphere including magnetic pores \citep{2011ApJ...729L..18M,2014A&A...563A..12D,2015A&A...579A..73M,2016ApJ...817...44F}, chromospheric network \citep{2007A&A...461L...1V,2010A&A...510A..41K,2014A&A...567A..62K}, coronal loops \citep{2015ApJ...811L..13W,2016NatPh..12..179J}, and polar plumes \citep{2011A&A...528L...4K,2014ApJ...793..117S}. Recent multi-wavelength observations of sunspots reveal that umbral flashes \citep{1969SoPh....7..351B,2003A&A...403..277R} and running penumbral waves \citep{1972SoPh...27...71G,2007ApJ...671.1005B,2014ApJ...791...61F} observed in the chromosphere, and quasi-periodic propagating disturbances observed in the coronal loops \citep{bc1999SoPh186,2000A&A...355L..23D,2009SSRv..149...65D}, are just different manifestations of the SMAWs that propagate from the photosphere through to the corona \citep{2012ApJ...757..160J,2015ApJ...812L..15K,2016ApJ...830L..17Z}. It has been proposed that the interaction of photospheric $p$-modes with magnetic fields generates different magnetohydrodynamic (MHD) waves \citep{1991LNP...388..121S, 1994ApJ...437..505C, 2015SSRv..190..103J}, of which SMAWs are guided upwards along the magnetic field lines \citep{2007AN....328..286C,2012ApJ...746...68K} allowing them to reach coronal heights \citep{2003ApJ...590..502D,2005ApJ...624L..61D}. Furthermore, it has been shown that oscillations with frequencies below the typical acoustic cutoff can also be channelled into the corona if the magnetic fields are inclined \citep{2005ApJ...624L..61D,2006ESASP.624E..15E}, or if the radiative losses are included \citep{2008ApJ...676L..85K}.

\begin{figure*}
\centering
\includegraphics[width=\textwidth, clip=true]{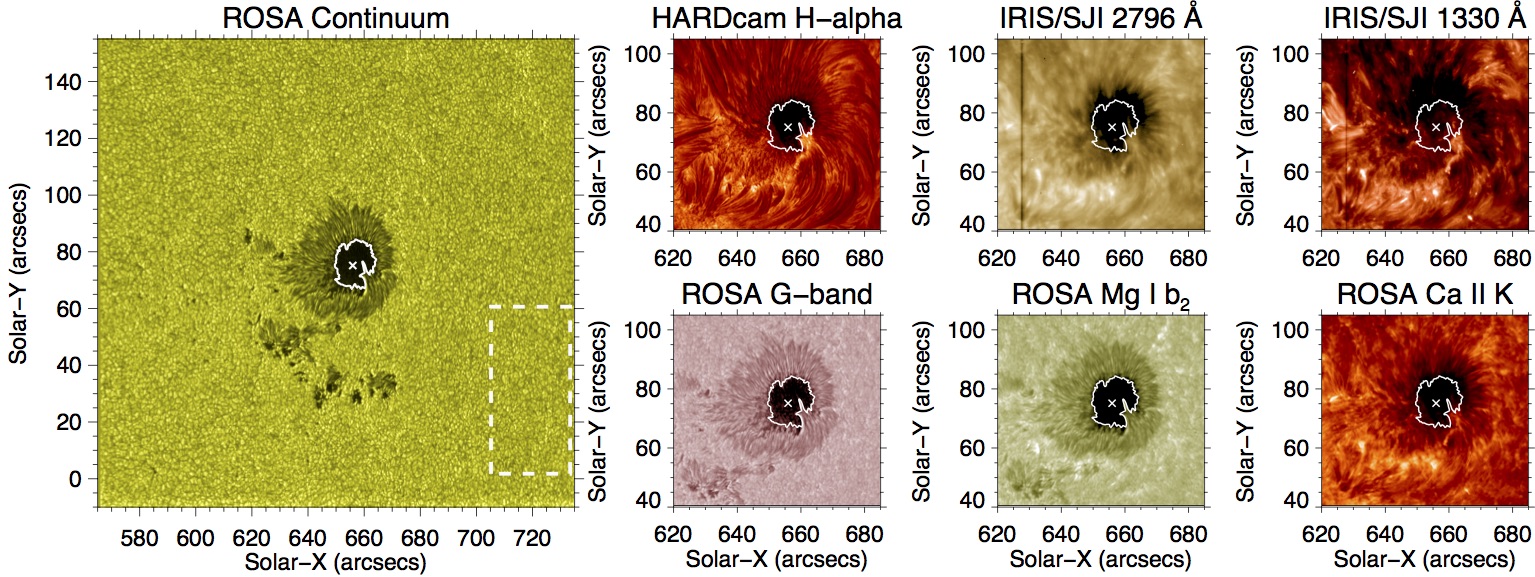} 
\caption{Sample images of a sunspot from AR 12149 in multiple wavelength channels, taken by ROSA, HARDcam, and IRIS/SJI. The image in ROSA continuum channel shows a larger field-of-view, while those in the rest display the closer vicinity of the sunspot. The white cross in the umbra locates the position of the umbral barycenter, while the solid white curve outlines the umbra-penumbra boundary. The white-dashed box in ROSA continuum image highlights the region selected to compute median intensity for ROSA/HARDcam data normalization (see text for details). The vertical black line in IRIS images shows the position of the IRIS slit outside the umbra.}
\label{fig1}
\end{figure*}

The available energy flux of SMAWs in the corona is on the order of a few hundred, to a few thousand erg{\,}cm$^{-2}${\,}s$^{-1}$ \citep{2006A&A...448..763M} which is not at all sufficient to maintain the corona at million-Kelvin temperatures \citep[$\sim$10$^{7}$ erg{\,}cm$^{-2}${\,}s$^{-1}$;][]{1977ARA&A..15..363W}. Nevertheless, the photospheric acoustic oscillations, particularly those at high frequencies, were thought to possess enough energy to replenish the chromospheric radiative losses. However, using the power spectra from quiet-Sun intensity fluctuations and one-dimensional numerical simulations, \citet{2005Natur.435..919F} have shown that the acoustic energy flux found in the 5--50{\,}mHz frequency range is substantially smaller than expected; about one-tenth of that required in the chromosphere. The authors employed data from the \textit{Transition Region and Coronal Explorer} \citep[TRACE;][]{1999SoPh..187..229H} in the 1700{\,}{\AA} and 1600{\,}{\AA} passbands to arrive at this conclusion. In contrast, three-dimensional numerical simulations by \citet{2007ASPC..368...93W} indicate the existence of sufficient energy flux from acoustic waves. Furthermore, \citet{2007ASPC..368...93W} demonstrate that the limited spatial resolution of TRACE observations underestimate the acoustic power by at least an order-of-magnitude, helping to explain the discrepancy. Subsequent observations with higher resolution have revealed larger, but still either insufficient \citep{2007PASJ...59S.663C} or barely comparable \citep{2009A&A...508..941B,2010ApJ...723L.134B} acoustic power required to dominate chromospheric heating. One must note, however, that the chromosphere, being magnetically rich with expanding magnetic fields, supports different MHD wave modes that are both compressible and incompressible \citep[e.g.,][]{2012NatCo...3E1315M,2013ApJ...779...82K, 2017arXiv170506282J}, which may make a significant contribution to localized heating.

In magnetized atmospheres, the energy contribution of acoustic oscillations (through SMAWs) to chromospheric heating is relatively less studied. Two-dimensional numerical simulations of small-scale network fields driven by transverse impulses reveal insufficient acoustic flux to balance chromospheric radiative losses \citep{2009A&A...508..951V}. Further two- and three-dimensional MHD simulations by \citet{2011ApJ...727...17F} and \citet{2012ApJ...755...18V} indicate variable acoustic energy flux in the upper atmospheric layers depending on the choice of the driver \citep[see also][]{2015ApJ...799....6M, 2015MNRAS.449.1679M}. In a sunspot umbra, using the root-mean-squared (rms) velocities of Na{\,}\textsc{i} D$_{1}$ and Na{\,}\textsc{i} D$_{2}$ lines, \citet{1981A&A...102..147K} estimate an outward energy flux of 5 $\times$ 10$^{4}$ erg{\,}cm$^{-2}${\,}s$^{-1}$, which is well below the chromospheric requirement of 2.6 $\times$ 10$^{6}$ erg{\,}cm$^{-2}${\,}s$^{-1}$ \citep{1981phss.conf..235A}. In a recent study, \citet{2011ApJ...735...65F} computed the acoustic energy flux as a function of atmospheric height using data-driven MHD simulations of a sunspot umbra. The authors found insufficient energy ($\approx$10$^{6}$ erg{\,}cm$^{-2}${\,}s$^{-1}$) even at the lowermost (photospheric) height investigated, which further decreases with height. \citet{2017ApJ...836...18C} calculate the average energy flux in three-minute oscillations over a sunspot umbra, observed in the Ni{\,}\textsc{i} 5436{\,}{\AA} line that forms 38{\,}km above photosphere, as 1.8 $\times$ 10$^{6}$ erg{\,}cm$^{-2}${\,}s$^{-1}$. This value is on the same order of that obtained by \citet{2011ApJ...735...65F} near the photosphere. On the contrary, based on the observational data of a sunspot umbra, \citet{2016ApJ...831...24K} estimate an energy flux of 2 $\times$ 10$^{7}$ erg{\,}cm$^{-2}${\,}s$^{-1}$ at the photospheric level, with 8.3 $\times$ 10$^{4}$ erg{\,}cm$^{-2}${\,}s$^{-1}$ at the lower transition region level in the 6--10{\,}mHz frequency band, implying a dissipation of sufficient energy to maintain the umbral chromosphere. The authors, however, add a caveat that the energies could be overestimated due to opacity effects.

In this article, we utilize high-resolution, high-cadence image sequences obtained simultaneously in multiple wavelengths, to study the damping of SMAWs in a sunspot umbra up to transition region heights. Representative power spectra of the sunspot umbra are generated across all channels to perform this in-depth study. We present the observational aspects of the data in section 2, followed by our analysis and results in section 3, and finally discuss our important interpretations in section 4.

\section{Observations}
The Dunn Solar Telescope (DST), situated in the Sacramento Peak mountains of New Mexico, was employed to obtain high-resolution images of active region NOAA 12149 at a very high cadence in five different wavelength channels using the Rapid Oscillations in the Solar Atmosphere \citep[ROSA;][]{2010SoPh..261..363J} and the Hydrogen-Alpha Rapid Dynamics camera \citep[HARDcam;][]{2012ApJ...757..160J}. The observations were carried out on 2014 August 30 starting from 14:37~UT for approximately 3 hours. Four identical ROSA cameras were used to capture images simultaneously in four wavelength channels corresponding to the blue continuum (4170{\,}\AA), G-band (4305.5{\,}\AA), Mg{\,}\textsc{i}{\,}b$_{2}$ (5172.7{\,}\AA), and Ca{\,}\textsc{ii}{\,}K line core (3933.7{\,}\AA), while the HARDcam instrument acquired images in the H$\alpha$ line core (6562.8{\,}\AA). The bandpass widths of each of these channels are 52{\,}{\AA}, 9.2{\,}{\AA}, 0.13{\,}{\AA}, 1{\,}{\AA}, and 0.25{\,}{\AA}, respectively. All of the images were processed following standard procedures. In addition to the application of high-order adaptive optics \citep{2004SPIE.5490...34R} during the observations, all of the data were subjected to speckle reconstruction algorithms \citep{2008A&A...488..375W} to improve the image quality. The final cadences of the data are 2.11{\,}s for the blue continuum, G-band and Ca{\,}\textsc{ii}{\,}K, 4.22{\,}s for the Mg{\,}\textsc{i}{\,}b$_{2}$ and 0.99{\,}s for the H$\alpha$ channels. 

\begin{table}
\begin{center}
\caption{Typical formation heights of different ROSA/HARDcam channels above photosphere.}
\label{tab1}
\begin{tabular}{l r}
\hline\hline
	Channel name					& Formation height  \\
\hline
	Blue continuum (4170{\,}{\AA})  	~~~~~			& 25~km \tablenotemark{a}\\
	G-band  						& 100~km \tablenotemark{a}\\
	Mg{\,}\textsc{i}{\,}b$_{2}$  	& 700~km \tablenotemark{b} \\
	Ca{\,}\textsc{ii}{\,}K 			& 1300~km \tablenotemark{c}\\
 	H$\alpha$ 						& 1500~km \tablenotemark{d}\\
\hline
\end{tabular}
\tablerefs{$^{a}$\citet{2012ApJ...746..183J}, $^{b}$\citet{1979A&A....74..273S}, $^{c}$\citet{1969SoPh...10...79B}, $^{d}$\citet{1981ApJS...45..635V}}
\end{center}
\end{table}

The data were subsequently normalized with a median intensity value obtained from a relatively quite region (outlined by a white-dashed box in Fig.~\ref{fig1}) in the field-of-view to compensate for brightness changes over time that result from changes in the elevation of the Sun. The atmospheric seeing conditions remained excellent throughout the time series, barring a few short-term local fluctuations that affected a relatively small number of images in each channel. These images were identified through the decreases in their contrast ratios with time, and were subsequently replaced through interpolation. Data from different channels were coaligned with respect to the blue continuum using a set of calibration images that were obtained immediately prior to the science observations. The spatial sampling of the ROSA data is $0{\,}.{\!\!}{\arcsec}18$ per pixel. HARDcam (H$\alpha$) images are observed at a better spatial resolution ($0{\,}.{\!\!}{\arcsec}09$ pixel$^{-1}$; as a result of the four-fold increase in pixel numbers), but for the analyses presented in this article, these data have been re-sampled to match the ROSA plate scale. The typical formation heights of individual ROSA and HARDcam channels are listed in Table~\ref{tab1}.

The Interface Region Imaging Spectrograph \citep[IRIS;][]{2014SoPh..289.2733D} also observed AR 12149 during the same timeframe. IRIS observations consist of sit-and-stare data lasting approximately 6 hours, commencing from 11:12~UT and lasting until 17:13~UT. The corresponding images taken by the slit-jaw imager (SJI) were obtained only in two channels, 2796{\,}{\AA} and 1330{\,}{\AA}. The spectrographic slit was pointed entirely outside the sunspot, therefore providing no associated spectra for the umbra under investigation. Thus, we consider only the SJI data, between 14:13 -- 17:13~UT, which has the maximum temporal overlap with the ROSA/HARDcam image sequences.  The 2796{\,}{\AA} and 1330{\,}{\AA} channels mainly capture the plasma present in the upper chromosphere and the transition region, respectively \citep{2014SoPh..289.2733D}. All of the data are processed to `level 2' standard, which incorporates dark current subtraction, flat-fielding and other necessary corrections, including the images being brought on to a common plate scale. Additionally, we also ensured each time sequence is properly coaligned with the first image using intensity cross correlations. The cadence of the data is approximately 18.8{\,}s and the spatial sampling is $\approx$$0{\,}.{\!\!}{\arcsec}166$ per pixel. Robust coalignment between the ROSA, HARDcam and IRIS channels has been achieved by cross-correlating the ROSA Ca{\,}\textsc{ii}{\,}K and IRIS 2796{\,}{\AA} images. Sample images for each of the channels are displayed in Fig.~\ref{fig1}. The ROSA blue continuum image displays the full ground-based field-of-view, while the rest display zoom-ins of the primary sunspot captured in each of the respective channels.\\

\begin{figure*} 
\centering
\includegraphics[width=0.85\textwidth, clip=true]{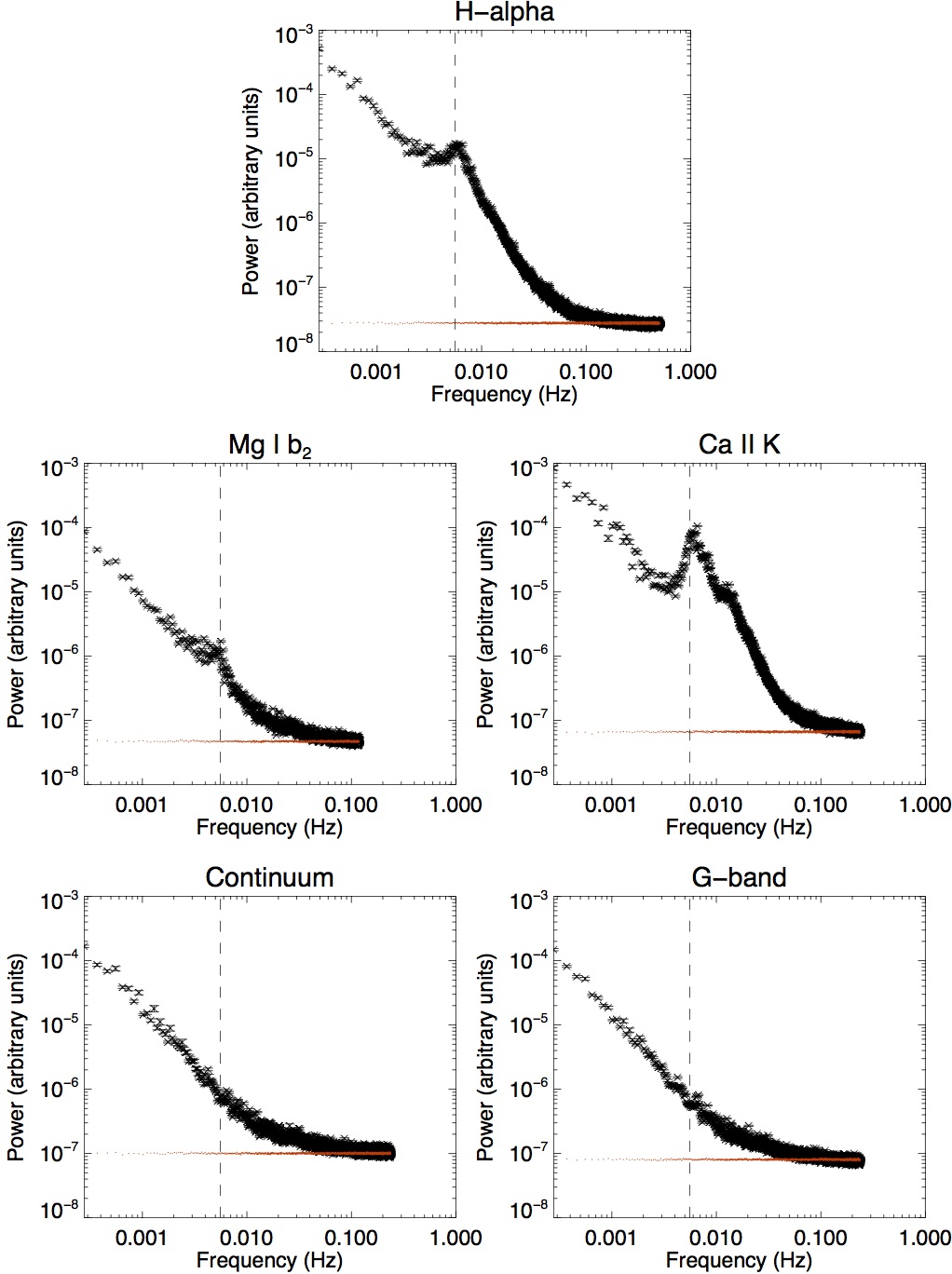}
\caption{Mean power spectra generated from intensity oscillations within the umbra, across multiple ROSA/HARDcam channels. The corresponding 1-sigma uncertainties on the values are shown as error bars which are on the order of the size of the symbol. The red dots display the level of photon noise present in the data, computed from artificially generated lightcurves that follow a Poisson distribution, with amplitudes equivalent to the square root of the mean observed intensity. The vertical dashed line denotes a frequency of 5.5{\,}mHz, where significant enhancements in power can be seen in chromospheric channels. Note the frequency range of the spectra differs in each of the channels due to differences in their respective cadences.}
\label{fig2}
\end{figure*}

\section{Analysis and Results}
Active region NOAA 12149 has a single, large, nearly-circular sunspot of negative polarity (Hale class $\beta/\beta$), surrounded by a few pore-like magnetic concentrations (of the same polarity) and some diffuse positive flux. The detailed structure of the sunspot and its vicinity across different ROSA, HARDcam and IRIS channels can be seen in Fig.~\ref{fig1}. Using the time-averaged blue continuum image, a boundary between the umbra and penumbra of the sunspot has been defined through intensity thresholding, calculated in relation to the median granulation intensity from the immediate surroundings. The location of the umbral barycenter in intensity is then computed from the pixel locations within this outer perimeter following the methods detailed by \citet{2013ApJ...779..168J}. The computed umbra-penumbra boundary and the umbral barycenter are marked by a white solid line and a white cross, respectively, in Fig.~\ref{fig1}. The excellent coalignment between different optical and UV channels may also be noted.

\subsection{Mean power spectra}
\label{mpspec}
A Fourier power spectrum of a time series reveals the oscillation frequencies manifesting within it as individual spikes in power. Time series from each pixel location within the umbra was subjected to Fourier analysis using the Fast Fourier Transform (FFT) method, with the corresponding frequencies, along with their respective power peaks, identified and noted. In addition to performing FFT analyses on a pixel-by-pixel basis, we also calculated a global average Fourier power spectrum for all pixels contained within the umbral perimeter using bootstrap method \citep{efron_1979}.

\begin{figure*}
\centering
\includegraphics[width=0.85\textwidth, clip=true]{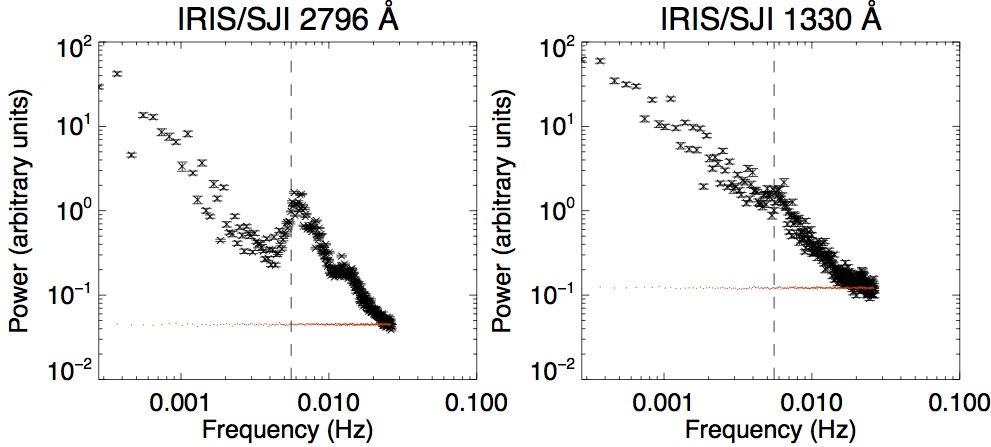}
\caption{Mean power spectra generated from intensity oscillations within the umbra using IRIS data. The corresponding 1-sigma uncertainties on the values are shown as error bars. The red dots indicate the respective levels of photon noise calculated in a similar way to that in Fig.~{\ref{fig2}}. The vertical dashed line marks the position of 5.5{\,}mHz. Note the extent of the spectra is lower than that in ROSA/HARDcam data (Fig.~{\ref{fig2}}) due to the lower temporal cadence of the IRIS data.}
\label{fig3}
\end{figure*}

The mean power spectra calculated following this method for all ROSA and HARDcam data are displayed in Fig.~\ref{fig2}, along with the respective 1-sigma uncertainties. As may be noted, the uncertainties are fairly small, smaller than the size of the symbol in most cases. The corresponding plots for the IRIS 2796{\,}{\AA} and 1330{\,}{\AA} channels are shown in Fig.~\ref{fig3}. Note that the highest frequency up to which the power spectra are plotted is different for each of the ROSA, HARDcam and IRIS channels due to differences in cadences of the respective datasets. The power in all channels decreases with frequency (implying a power-law dependence), albeit with an intermediate bump visible in some channels and a plateau of power at higher frequencies. 

The flattening of power at higher frequencies is due to the white noise that dominates the signal at those frequencies. Assuming enough photon statistics\footnote{i.e., the signal is well above the detector background noise characteristics such as the dark noise etc.} in the umbra, the white noise is primarily composed of photon noise. The photon noise, by nature, follows a Poisson distribution, and has an amplitude proportional to the square root of the signal. To estimate the level of this noise present in the data, for each umbral pixel, we generate a random lightcurve following a Poisson distribution, with an amplitude equivalent to the square root of the mean intensity at that pixel. Using these artificially generated noise lightcurves, we compute the mean power spectrum in an identical way to that of the original data. Hence, the modeled noise power spectra, for each channel after appropriate scaling, are shown as red dots in Figs.~\ref{fig2} \& \ref{fig3}.

\begin{figure*}
\centering
\includegraphics[width=0.85\textwidth, clip=true]{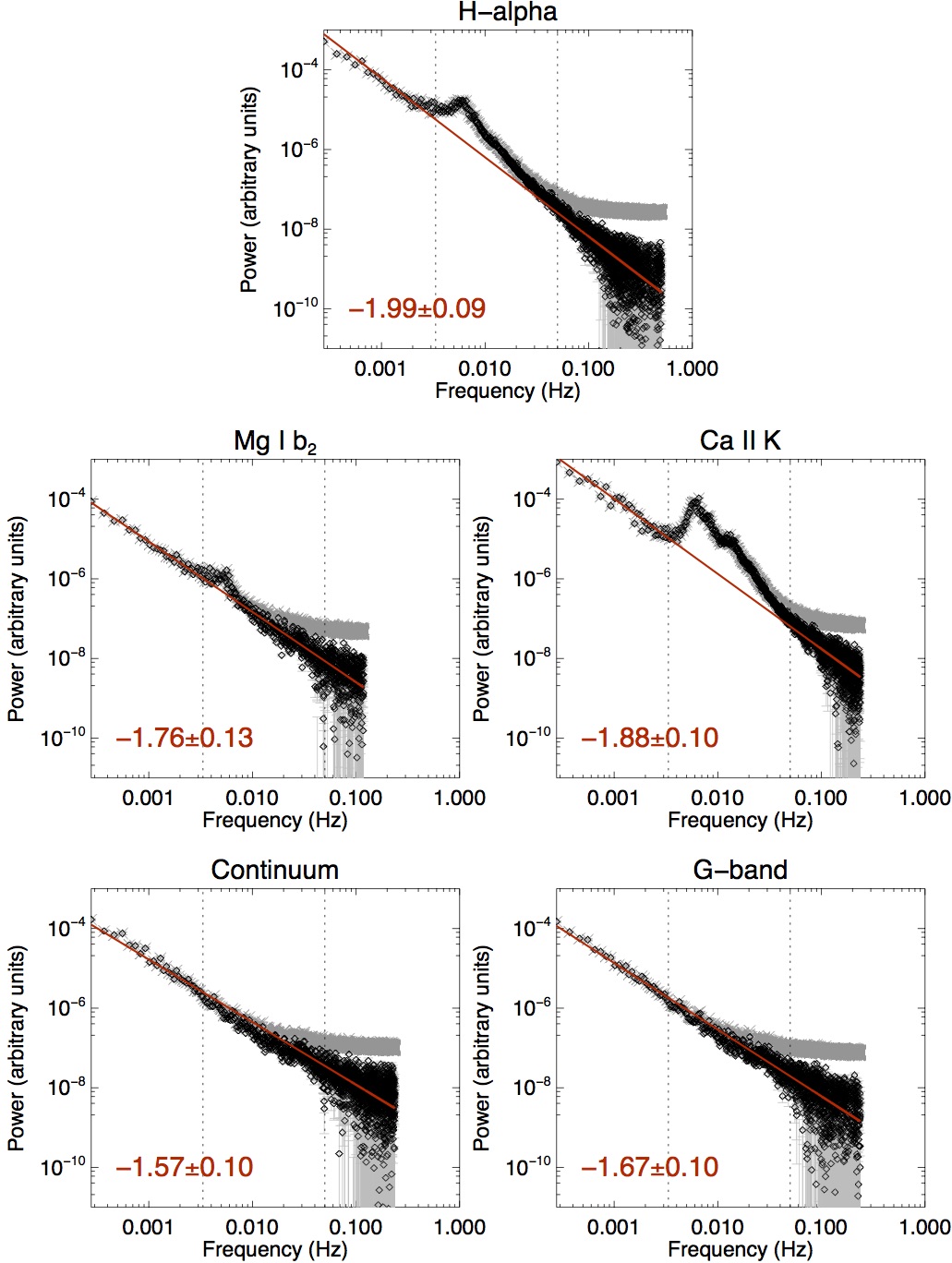}
\caption{The resultant power spectra in ROSA/HARDcam channels, following the subtraction of photon noise from each of the mean power spectra. The vertical bars in grey represent the corresponding uncertainties. The original mean power spectra are also shown in grey for comparison. Red solid lines represent the best linear fits to the data using a logarithmic scale, highlighting a clear power-law dependence. The respective power-law indices, as obtained from the slopes, are also listed in each plot. The data between the vertical dashed lines, positioned at 3.3 and 50{\,}mHz, are ignored while performing the linear fits.}
\label{fig4}
\end{figure*}

\begin{figure*}
\centering
\includegraphics[width=0.85\textwidth, clip=true]{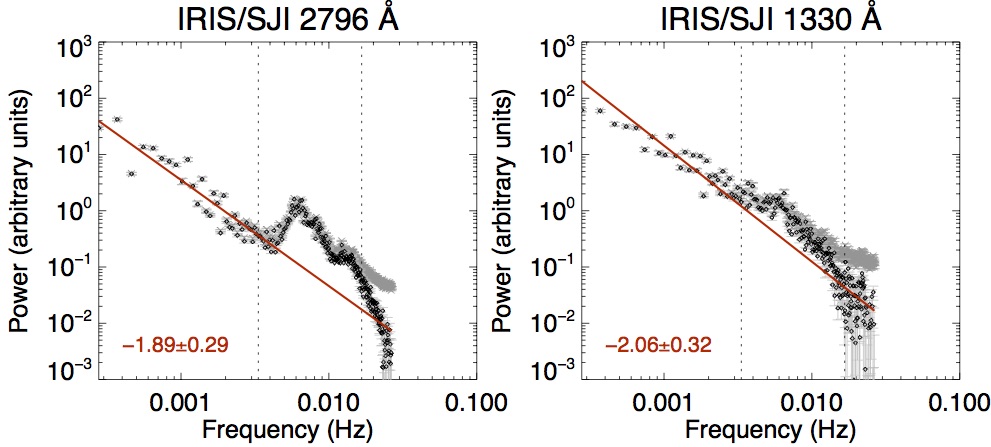}
\caption{Same as Fig.~\ref{fig4}, but here for IRIS data. The data between 3.3 and 16.6{\,}mHz are ignored during fitting, as denoted by the vertical dashed lines.}
\label{fig5}
\end{figure*}

The computed noise levels are subtracted from the individual power spectra, with the resultant power spectra for all ROSA, HARDcam and IRIS channels displayed in Figs.~\ref{fig4} \& \ref{fig5}. In these figures, the vertical bars in grey represent the corresponding uncertainties which are substantially larger at high frequencies due to the lower power values obtained after white noise subtraction. Also, the errors appear asymmetric due to the log-scale employed. It may be noted that the applied correction for white noise is similar to the methods adopted by \citet{2005Natur.435..919F} and \citet{2011ApJ...743L..24L}. The original power spectra are also shown in these plots, in grey, for comparison. It seems that, following the subtraction of white noise, the same power-law dependence continues to be present at higher frequencies. In order to quantify the exact dependence, each power spectrum, in a logarithmic scale, was fitted linearly following least-squares minimization \citep{2009ASPC..411..251M}. Red solid lines in Figs.~\ref{fig4} \& \ref{fig5} indicate the best fits to the data. The corresponding slopes, which represent the power-law indices in a linear scale, are also listed in each plot. The data in the intermediate peaks, between 3.3 -- 50~mHz (300 -- 20~s; denoted by vertical dotted lines in the figure) are ignored while fitting the power spectra for ROSA and HARDcam channels. This range is restricted to 3.3 -- 16.6 mHz (300 -- 60 s) for IRIS data, due to their lower cadence. Also, since there are substantially larger number of data points at high frequencies (righthand side of the peak) than those at the low frequencies (lefthand side of the peak), the least-squares fit would be strongly biased towards the high-frequency data which is not desirable. In order to avoid this, we used differential weights\footnote{The weights for all the low-frequency data ($<$3.3~mHz) were set to unity and for all the high-frequency data ($>$ 50~mHz for ROSA and $>$16.6~mHz for IRIS data) were set to 0.1, while those for the intermediate frequencies were set to zero.} for the low- and the high-frequency data while fitting. The resultant fitted slopes, as shown in Figs.~\ref{fig4} and \ref{fig5}, appear to increase with solar atmospheric height.

The intermediate bumps in the power spectra peak near 5.5{\,}mHz ($\approx$180{\,}s; marked by a vertical dashed line in Figs.~\ref{fig2} \& \ref{fig3}), which is a characteristic frequency of SMAWs often observed in the umbral chromosphere. The photospheric umbra usually displays oscillations at a frequency $\approx$3.3{\,}mHz. Umbral oscillations at these frequencies are believed to be connected to the photospheric $p$-modes \citep[e.g.,][]{2015ApJ...812L..15K,2016ApJ...830L..17Z}. Several theories also exist to explain the transition of the characteristic frequency from 3.3{\,}mHz in the photosphere to 5.5{\,}mHz in the chromosphere \citep{1983SoPh...82..369Z,1991A&A...250..235F}. However, oscillation amplitudes in the photosphere are usually very low, which are further minimized by the opacity effects present in intensity measurements \citep[e.g.,][]{2015LRSP...12....6K}. This is perhaps the reason why no bumps are observed in the blue continuum and G-band power spectra. It may be noted that the level of enhancement in the bump is variable across the different imaging channels. This is due to the varying amplitude of oscillations as the waves propagate through different atmospheric layers. However, a direct comparison of the power in different channels is not particularly trivial since it depends on several physical and instrumental factors, including the spectral region, exposure time, filter bandwidth, etc. Therefore, the power spectra in each channel are normalized to their respective power-law fits, which provides a relative indication of the amplitude of oscillations present in that atmospheric layer, and thus may be safely compared across different channels to gain additional physical insight.

\begin{figure*}
\centering
\includegraphics[width=\textwidth, clip=true]{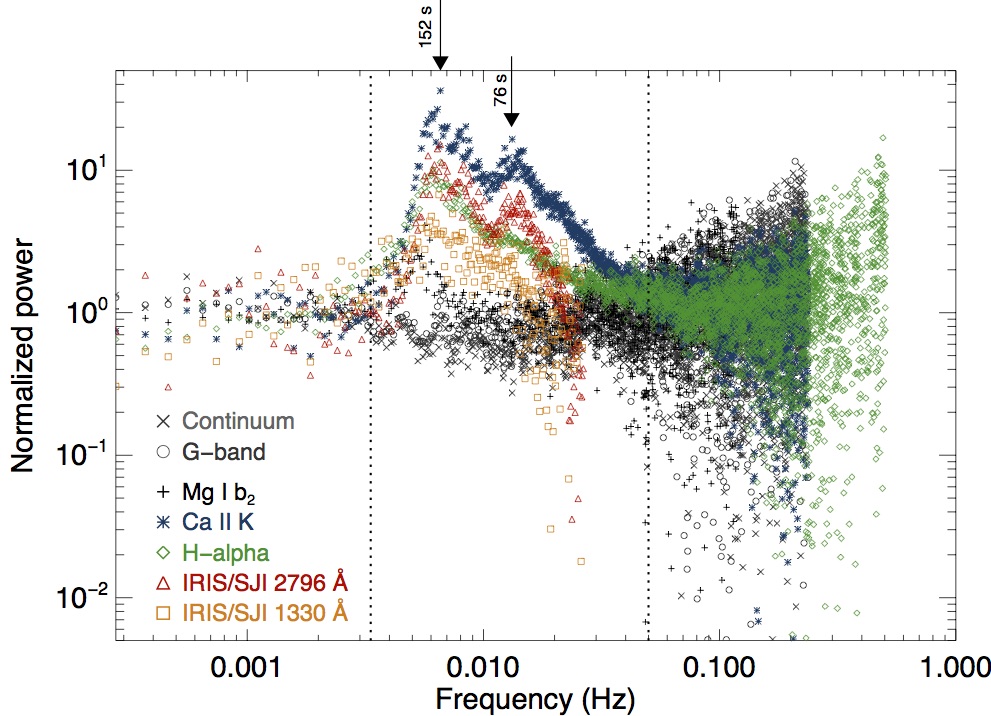}
\caption{Normalized power in various ROSA, HARDcam and IRIS channels, obtained by dividing the individual power spectra with their respective power-law fits. Different symbols and colors are used to denote the spectra described in the plot legend. The uncertainties on these data are similar to those shown in Figs.~\ref{fig4} and \ref{fig5} with larger values at high frequencies. The vertical dotted lines denote the positions of 3.3{\,}mHz and 50{\,}mHz frequencies. The arrows at 6.57{\,}mHz (152{\,}s) and 13.1{\,}mHz (76{\,}s) show the locations of the two harmonic peaks observed in the Ca{\,}\textsc{ii}{\,}K and IRIS 2796{\,}{\AA} channels. The power in Mg{\,}\textsc{i}{\,}b$_{2}$ peaks at 5.65~mHz ($\approx$177{\,}s). Note the different strengths of the peaks found in different channels, which indicates varying oscillation amplitudes.}
\label{fig6}
\end{figure*}

Fig.~\ref{fig6} displays the normalized power spectra for various ROSA, HARDcam and IRIS channels. Different colors and symbols are used to denote the data from different channels. The uncertainties on these data are similar to those displayed in Figs.~\ref{fig4} and \ref{fig5}, but are not shown here to avoid cluttering from larger values at high frequencies. The vertical dotted lines mark the locations of 3.3~mHz (300~s) and 50~mHz (20~s). As can be seen, there is no obvious peak in the blue continuum and G-band channels. The power present in the rest of the channels peaks at $\approx$6.57~mHz (152~s),\footnote{In IRIS 1330{\,}{\AA} channel, the power peaks at 6.48{\,}mHz (154~s).} barring the Mg{\,}\textsc{i}{\,}b$_{2}$ channel which peaks at $\approx$5.65~mHz (177~s). Besides the main peak, two additional peaks are visible in the ROSA Ca{\,}\textsc{ii}{\,}K and IRIS 2796{\,}{\AA} channels at $\approx$8.3~mHz (120~s) and $\approx$13.1~mHz (76~s) frequencies, of which the latter is more prominent. 

\begin{figure*}
\centering
\includegraphics[width=\textwidth, clip=true]{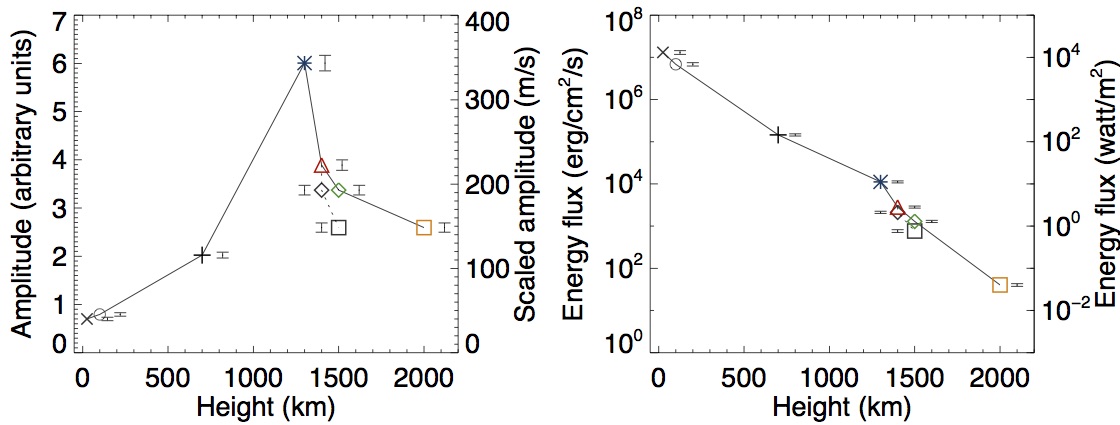}
\caption{\textit{Left}: Relative amplitudes of 5.5{\,}mHz (three-minute) oscillations at different heights, derived from the normalized peak power in various ROSA, HARDcam and IRIS channels. Different colors and symbols are used to represent different channels following the same notation as in Fig.~\ref{fig6}. The grey diamond and square represent the data from the H$\alpha$ and IRIS 1330{\,}{\AA} channels, respectively, assuming alternative formation heights, as obtained from the phase-difference spectra (see Section~\ref{phd_spec} for details). The axis on the right shows the amplitude values after scaling the blue continuum Fourier amplitudes to 40{\,}m{\,}s$^{-1}$. \textit{Right}: Corresponding energy flux obtained by using temperatures and densities from the \citet{1986ApJ...306..284M} umbral ``M'' sunspot model. The axis on the right shows the same energy flux values in S.I. units. Respective uncertainties on the amplitude and the energy flux values are shown as vertical bars, plotted adjacent to the data for clarity.}
\label{fig7}
\end{figure*}

The peak power appears to increase from Mg{\,}\textsc{i}{\,}b$_{2}$ to Ca{\,}\textsc{ii}{\,}K, before further decreasing in the neighboring H$\alpha$ and IRIS channels. This implies an increase and decrease of the oscillation amplitudes across these channels. Considering the square root of peak power in each channel as a measure of oscillation amplitude, we plot the variation of amplitude with atmospheric height in the left panel of Fig.~\ref{fig7}. Colors and symbols used to represent data from different channels, are same as that in Fig.~\ref{fig6}. The corresponding uncertainties are shown as vertical bars plotted adjacent to the data for clarity. The power at 5.55~mHz ($\approx$180~s) is used to calculate the blue continuum and G-band amplitudes. As listed in Table~\ref{tab1}, the formation heights chosen for the blue continuum, G-band, Mg{\,}\textsc{i}{\,}b$_{2}$, Ca{\,}\textsc{ii}{\,}K and H$\alpha$ are 25~km \citep{2012ApJ...746..183J}, 100~km \citep{2012ApJ...746..183J}, 700~km \citep{1979A&A....74..273S}, 1300~km \citep{1969SoPh...10...79B} and 1500~km \citep{1981ApJS...45..635V} above the photosphere, respectively. The IRIS 2796{\,}{\AA} channel, with a bandpass of 4{\,}{\AA}, is sensitive to the plasma present in the upper chromosphere, while the 1330{\,}{\AA} channel, with a bandpass of 55{\,}{\AA}, is sensitive to transition region plasma \citep{2014SoPh..289.2733D}. This is also evident from the structures visible in the IRIS images (Fig.~\ref{fig1}). Taking the close resemblance between Ca{\,}\textsc{ii}{\,}K and IRIS 2796{\,}{\AA} images into account, the formation height of the 2796{\,}{\AA} channel is chosen as 1400{\,}km, while the 1330{\,}{\AA} channel is deemed to be representative of 2000{\,}km above the photosphere (consistent with typical transition region heights). Alternative formation heights that may be possible for the H$\alpha$ and IRIS 1330{\,}{\AA} channels, as inferred from the phase difference spectra (see Section~\ref{phd_spec}), are used to mark the grey diamond and square, respectively, in Fig.~{\ref{fig7}}. The corresponding errors are shown on lefthand side for these data. We must emphasize that the plotted amplitudes obtained from the normalized power do not in any way constitute absolute values, but instead provide a representative comparison between the strengths of the oscillations observed at different atmospheric heights.

We also computed the energy flux ($F$) following WKB approximation, using $F = \rho \langle \delta v^2\rangle  c_s$, where $\rho$, $\delta v$, and $c_s$, are the mass density, velocity amplitude, and sound speed respectively. The sound speed $c_s$ is related to the temperature $T$, as $c_s$=$\sqrt{\gamma RT/ \mu}$, where $\gamma$ is the polytropic index, $R$ is the gas constant, and $\mu$ is the mean molecular weight. We consider $\gamma$=5/3, $R$=8.314$\times 10^{7}$ erg{\,}K$^{-1}${\,}mol$^{-1}$, and $\mu$=0.61 \citep{1993str..book.....M}, in these calculations. The required temperature and density values corresponding to the observational formation heights were acquired through spline interpolation of the values extracted from the umbral core ``M'' model of \citet{1986ApJ...306..284M}. The velocity amplitudes are obtained by scaling the ROSA blue continuum Fourier amplitude to 40{\,}m{\,}s$^{-1}$. This corresponds to the rms velocity amplitude (integrated in the 5--8~mHz band) averaged over two sunspot umbrae observed by \citet{1985ApJ...294..682L} in the Ti~\textsc{i} 6304{\,}{\AA} line that forms $\approx$40{\,}km above the photosphere \citep{1984ssdp.conf..141A}, close to the formation height of the ROSA blue continuum channel. Furthermore, from a compilation of similar results from several authors it is shown that the rms oscillatory amplitudes in the 5--8~mHz (2--3~min) band in the umbral photosphere falls within the range of 25--50~m{\,}s$^{-1}$ \citep[Table I;][]{1992ASIC..375.....T}. Also, more recently, \citet{2017ApJ...836...18C} report similar amplitudes (33~m{\,}s$^{-1}$) for three-minute oscillations over a sunspot umbra using observations in the Ni{\,}\textsc{i} 5436{\,}{\AA} line that forms $\approx$38{\,}km above photosphere. The axis on the righthand side of left panel in Fig.~{\ref{fig7}} shows the scaled amplitudes in m{\,}s$^{-1}$. The computed energy flux values are shown in the right panel of Fig.~\ref{fig7}, with an additional axis on the right to show the energy flux in S.I. units (W{\,}m$^{-2}$). The grey diamond and square indicate the energy flux in the H$\alpha$ and IRIS 1330{\,}{\AA} channels, respectively, assuming alternative formation heights for the data based on phase difference analysis (see Section~\ref{phd_spec}). The error bars adjacent to the data represent the corresponding uncertainties. As can be seen, the energy flux gradually decreases, even when the oscillation amplitude is increasing, suggesting the damping of SMAWs across these atmospheric layers. It may be noted that modifying the velocity amplitudes in the photosphere will change the energy flux by an order-of-magnitude or more. However, the decreasing trend in energy flux, where our main emphasis lies, remains unchanged.

\begin{figure}
\centering
\includegraphics[width=\columnwidth, clip=true]{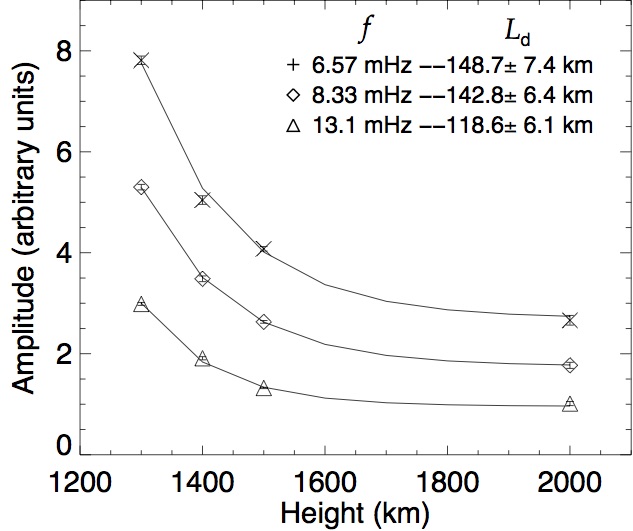}
\caption{Oscillation amplitudes, along with the corresponding uncertainties, at three different frequencies obtained from the original power spectra for Ca{\,}\textsc{ii}{\,}K, IRIS 2796{\,}{\AA}, H$\alpha$, and IRIS 1330{\,}{\AA} channels. Each of the spectra were normalized with their respective power at 3 mHz to enable comparison across multiple channels. The power integrated over a 1 mHz band chosen around each frequency, is used to compute the amplitudes. Solid lines represent exponential fits to the data. The damping lengths ($L_{d}$) obtained from the fitted values along with the respective central frequencies ($f$) are listed in the plot.}
\label{fig8}
\end{figure}

Slow magnetoacoustic waves have been reported to undergo frequency-dependent damping in the solar corona \citep{2002ApJ...580L..85O,2014ApJ...789..118K}. In order to explore if such behavior is also exhibited by these waves in the sub-coronal layers, we computed oscillation amplitudes at three frequencies coincident with the identified peaks at 6.57, 8.33, and 13.1 mHz. The original power spectra (after correction for white noise) were used to calculate the amplitudes to avoid potential influences of the varying power-law slopes on the frequency-dependence. The individual power spectra were, however, normalized with their respective power at 3 mHz to enable comparison across different channels. The power integrated in a 1 mHz band around each of the frequency peaks (0.5 mHz on either side of the peak) is used to evaluate the amplitudes. The data in Ca{\,}\textsc{ii}{\,}K, IRIS 2796{\,}{\AA}, H$\alpha$, and IRIS 1330{\,}{\AA} channels synonymous with the declining phase of amplitudes, are only considered. Fig.~\ref{fig8} displays the computed amplitudes as a function of formation height of the channels along with the respective uncertainties. Solid lines represent the best fits to the data following a function $A=A_{0} e^{-\frac{h}{L_{d}}}+C$, where $A$ is the oscillation amplitude, $h$ is the formation height, $L_{d}$ is the damping length, and $A_{0}$ and $C$ are appropriate constants. The exponential decay function appears to be consistent with the data. Damping lengths obtained from the fits, at each frequency, $f$, are listed in the plot, which indicate frequency-dependent damping of the waves with stronger damping found at higher frequencies.\\

\subsection{Phase difference spectra}
\label{phd_spec}
The phase difference between oscillations at different heights can be used to determine whether the waves are upward/downward propagating or standing \citep[e.g.,][]{2006ApJ...640.1153C}. We computed phase differences at all frequencies between the channel pairs (Mg{\,}\textsc{i}{\,}b$_{2}$, Ca{\,}\textsc{ii}{\,}K), (Mg{\,}\textsc{i}{\,}b$_{2}$, H$\alpha$), (Ca{\,}\textsc{ii}{\,}K, H$\alpha$) and (IRIS 2796{\,}{\AA}, 1330{\,}{\AA}), using their respective cross-power spectra. Data from individual pixel locations were directly matched in these calculations, assuming negligible magnetic field expansion between the formation heights of channels in each pair. This assumption is perhaps reasonable within a sunspot umbra where the magnetic field is mostly vertical. Also, the Ca{\,}\textsc{ii}{\,}K and H$\alpha$ data were re-sampled to match the lower temporal cadence of the Mg{\,}\textsc{i}{\,}b$_{2}$ channel, which made the comparison between these channels more straightforward.

\begin{figure*}
\centering
\includegraphics[width=0.85\textwidth, clip=true]{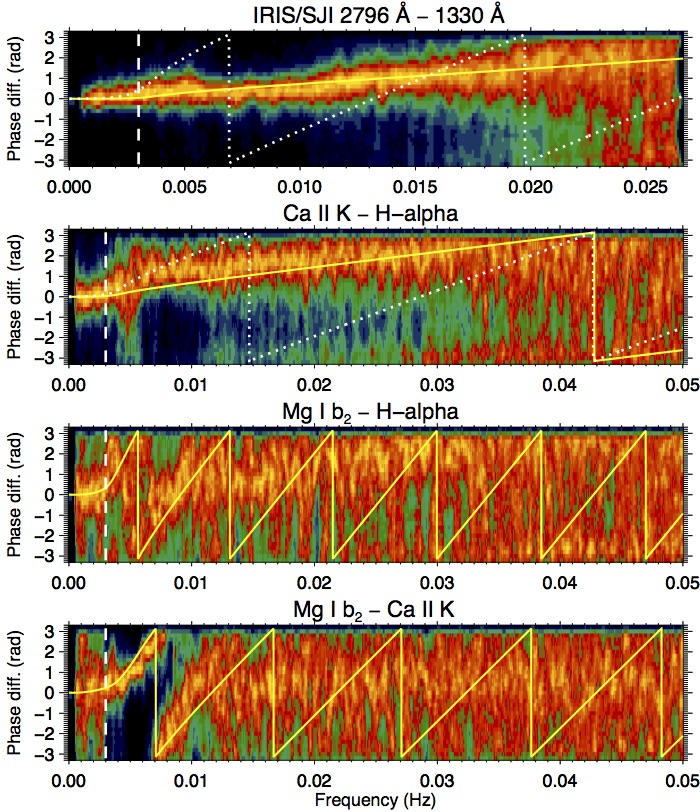}
\caption{Phase-difference spectra for different channel pairs computed for all the pixel locations within the umbra. The vertical axis shows the phase difference in radians, while the color scale indicates the relative occurrences of a particular phases, with red/yellow representing high and blue/green representing low values. The yellow solid curves denote the corresponding best-fit theoretical phase difference spectra consistent with the work of \citet{2006ApJ...640.1153C}. The white-dotted curves in the top two panels show the expected theoretical phases using typically published formation heights for the individual channels. The vertical white-dashed lines marks the positions of the 3{\,}mHz frequency that distinguishes evanescent and propagating waves.}
\label{fig9}
\end{figure*}

The resulting phase difference spectra, incorporating the data from the entire umbra, are plotted in Fig.~\ref{fig9} for all of the channel pairs. At each frequency, the phase difference values across the umbral pixels are grouped into histograms that constitute the color scale employed in each plot. Therefore, the color only indicates the predominance of one value over the other, with red/yellow representing more frequent and blue/green representing less frequent values. These diagrams are similar to those obtained by \citet{2010ApJ...722..131F,2011ApJ...735...65F}. The phase difference appears to repeat cyclically between $-\pi$ and $+\pi$ due to the 2$\pi$ indetermination in the computed phases previously noted by \citet{2006ApJ...640.1153C}. The values are plotted only up to 50~mHz as the photon noise dominates over the signal beyond this range (see Fig.~\ref{fig2}). Note that the IRIS data are plotted only up to $\approx$27{\,}mHz due to their lower cadence. Following \citet{2006ApJ...640.1153C}, we generated a theoretical phase difference curve (yellow solid lines in Fig.~\ref{fig9}) for each channel pair. Wave propagation in a stratified isothermal atmosphere with radiative cooling \citep{2006ApJ...640.1153C} is considered. However, to accommodate for the varying temperatures along the wave propagation path, phase differences are calculated in steps of 100~km (within which the isothermal approximation is employed) between the estimated formation heights of the respective channels, which are then integrated over the full path to get the final phase difference curves for a channel pair. The required temperature values at each step are obtained from the umbral core ``M'' model of \citet{1986ApJ...306..284M} through spline interpolation. The radiative cooling time is kept constant at 45{\,}s, 20{\,}s, 15{\,}s and 10{\,}s for the channel pairs (Mg{\,}\textsc{i}{\,}b$_{2}$, Ca{\,}\textsc{ii}{\,}K), (Mg{\,}\textsc{i}{\,}b$_{2}$, H$\alpha$), (Ca{\,}\textsc{ii}{\,}K, H$\alpha$) and (IRIS 2796{\,}{\AA}, 1330{\,}{\AA}), respectively. These values are approximately on the same order of those used in previous calculations \citep{2006ApJ...640.1153C,2007ApJ...671.1005B,2010ApJ...722..131F}.

As can be seen from the phase difference spectra for the (Mg{\,}\textsc{i}{\,}b$_{2}$, Ca{\,}\textsc{ii}{\,}K) and (Mg{\,}\textsc{i}{\,}b$_{2}$, H$\alpha$) pairs, the phase difference initially stays near zero up to about 3{\,}mHz (marked by a white dashed line in Fig.~\ref{fig9}), before increasing linearly beyond. This implies oscillations with frequencies above 3{\,}mHz are propagating upwards, while those below are evanescent between the two layers. Previous studies on sunspot umbrae have shown a similar cutoff frequency at about 4{\,}mHz \citep{2006ApJ...640.1153C, 2010ApJ...722..131F}, but as demonstrated by \citet{2009ApJ...692.1211C}, the exact value depends on the physical conditions of the structure. Theoretical phase difference curves are computed assuming the formation heights of 700{\,}km for Mg{\,}\textsc{i}{\,}b$_{2}$, 1300{\,}km for Ca{\,}\textsc{ii}{\,}K and 1500{\,}km for H$\alpha$ channels (see Table~\ref{tab1}). The theoretical curves show a reasonably good fit until about 13{\,}mHz, beyond which the observed phase differences appear noisy. It must be noted that the phase difference spectra represents values from individual pixel locations, unlike the power spectra shown in Figs.~\ref{fig2}--\ref{fig5} that display averages across the entire sunspot umbra. This perhaps explains why the phase differences at higher frequencies appear noisier, even though the oscillation power is well above the background noise level. This could also be partly related to the lower cadence of the Mg{\,}\textsc{i}{\,}b$_{2}$ observations, resulting in more significant photon noise at relatively lower frequencies compared to Ca{\,}\textsc{ii}{\,}K and H$\alpha$ channels (see Fig.~\ref{fig4}).

The phase difference spectra for the (Ca{\,}\textsc{ii}{\,}K, H$\alpha$) and (IRIS 2796{\,}{\AA}, 1330{\,}{\AA}) pairs are shown in the top two panels of Fig.~\ref{fig9}. Here, the theoretical phase difference curves assume formation heights of 1300{\,}km, 1400{\,}km, 1500{\,}km and 2000{\,}km for the Ca{\,}\textsc{ii}{\,}K, IRIS 2796{\,}{\AA}, H$\alpha$ and IRIS 1330{\,}{\AA} channels, respectively, which are shown as white dotted lines. Note the oscillation amplitudes in these channels are fairly high when compared to those found in the Mg{\,}\textsc{i}{\,}b$_{2}$ observations (see Figs.~\ref{fig6} \& \ref{fig7}), which makes these phase difference spectra reliable up to frequencies as high as 13{\,}mHz, and possibly beyond. Hence, the significant departure of the observed phase differences from the theoretical values might indicate the assumed formation heights are wrong. Indeed, the yellow solid lines, which seem to agree well with the observations, correspond to a formation height difference of $<$100{\,}km between each channel. Nevertheless, the positive phase difference values, combined with the already inferred upward propagation between Mg{\,}\textsc{i}{\,}b$_{2}$ and H$\alpha$ heights, implies a higher formation height for H$\alpha$ and IRIS 1330{\,}{\AA} when compared to Ca{\,}\textsc{ii}{\,}K and IRIS 2796{\,}{\AA}, respectively. While we acknowledge the fact that it is not trivial to assign a single formation height to any of these channels, the remarkable differences in the visible structures of the sunspot, between the Ca{\,}\textsc{ii}{\,}K and H$\alpha$ channels and between the IRIS 2796{\,}{\AA} and 1330{\,}{\AA} channels, suggest the difference in formation heights could be larger than 100 km. However, the relatively broad filter bandpasses of the Ca{\,}{\sc{ii}}{\,}K and IRIS SJI channels may also contribute to these uncertainties, since different regions of the solar atmosphere may dominate their respective contribution functions in drastically different ways. Indeed, it is possible that the observed formation height differences may be much smaller within the umbra, where the plasma is inherently cooler with significantly reduced opacities, when compared to other plage and quiet Sun locations. On the other hand, SMAWs propagating above the Ca{\,}\textsc{ii}{\,}K height may be predominantly nonlinear, and since the phase speed of nonlinear waves is larger than their linear counterparts, the difference in formation heights obtained from the linear wave theory may be substantially underestimated \citep{2010ApJ...722..131F}. Also, one must note that the phase difference spectra generated from the intensity oscillations alone is often difficult to interpret, unlike those from the velocity oscillations, since the phase relation between intensity and velocity is a complex function of frequency and radiative losses \citep{1984oup..book.....M,1989A&A...213..423D} which perhaps could partially explain the discrepancy.\\

\subsection{Radial variation}
In section~\ref{mpspec}, we addressed the mean power spectra computed from all pixel locations within the umbra. However, it is important to know whether the power spectra remain the same throughout the entire umbra. In order to study this, we implement the methods presented by \citet{2016NatPh..12..179J} and generate mean power spectra from a series of expanding annuli centered on the umbral barycenter. The width of each annulus is kept at 1~pixel and the radius is varied from 0 to the farthest point on the umbra-penumbra boundary. A mean power spectrum is calculated from all pixel locations falling within an annulus following the same procedures described in Section~\ref{mpspec}. For the outer annuli, any pixel locations falling outside the umbral penumbral boundary are ignored in the subsequent calculations.

\begin{figure*}
\centering
\includegraphics[width=0.85\textwidth, clip=true]{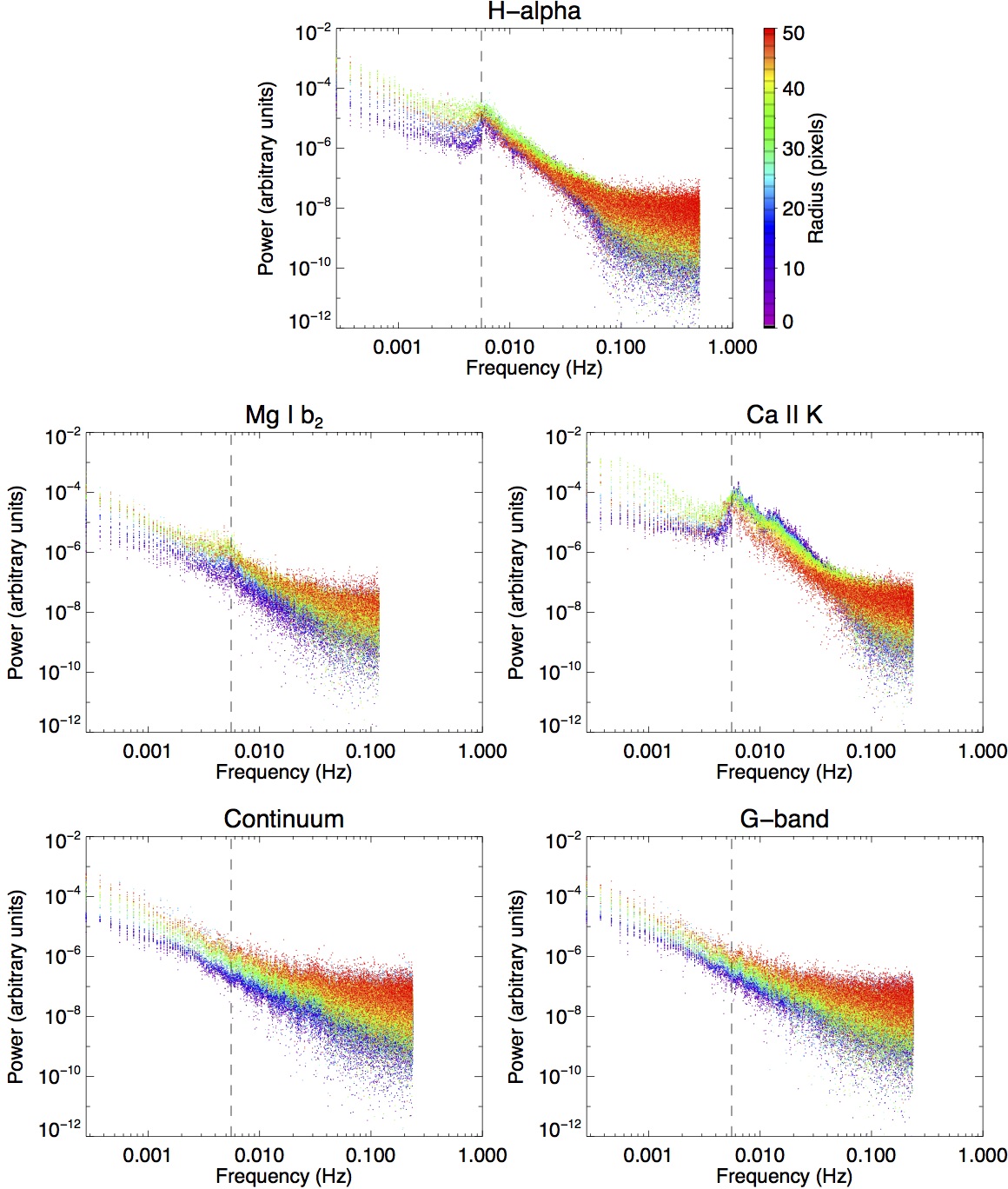}
\caption{Mean power spectra extracted from an expanding series of annuli spanning from the umbral barycenter to the umbra-penumbra boundary. The spectra from different annuli are colored in accordance with their radii, as indicated by the color scale shown adjacent to the top panel. Power due to photon noise is subtracted from the individual spectra before plotting. The vertical dashed line marks the position of 5.5{\,}mHz where strong chromospheric peaks are identified.}
\label{fig10}
\end{figure*}

\begin{figure*}
\centering
\includegraphics[width=\textwidth, clip=true]{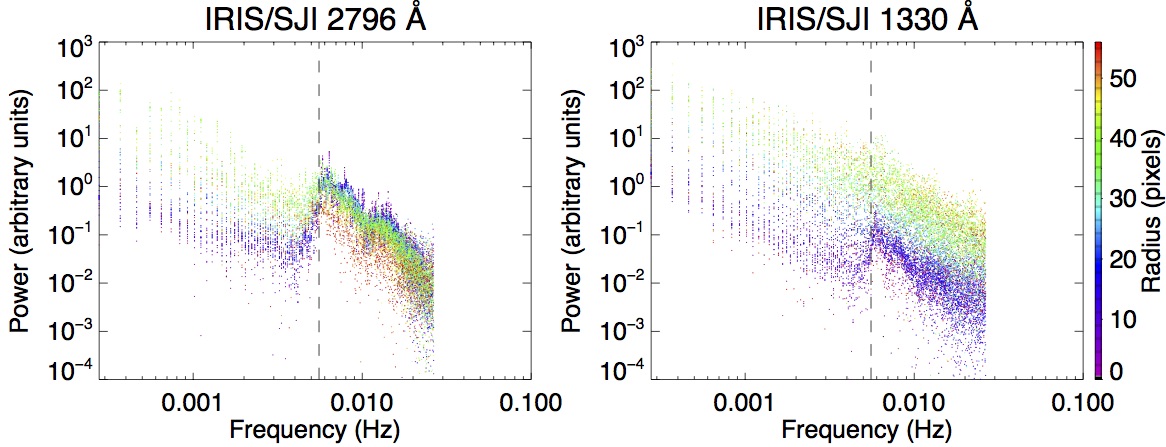}
\caption{Same as Fig.~\ref{fig10}, but for IRIS data. Note that the maximum radius here is larger than that in the ROSA/HARDcam data due to finer spatial sampling.}
\label{fig11}
\end{figure*}

The obtained power spectra for the ROSA/ HARDcam and IRIS data are shown in Figs.~\ref{fig10} \& \ref{fig11}, respectively. The color scale in each of these plots documents the radius of the annulus (in pixels) from which the power spectra is generated. Note that the maximum radius in the IRIS data is slightly larger due to its finer spatial sampling. As described in Section~\ref{mpspec}, the corresponding photon noise power has been calculated and subtracted from the individual spectra before plotting. Some flattening is still evident at high frequencies, which may be a result of limited number statistics pertaining to the 1-pixel wide annuli employed. Nevertheless, the overall frequency dependence appears to be roughly the same as in Figs.~\ref{fig4} \& \ref{fig5}. The corresponding uncertainties are also similar but not shown in these plots, to avoid cluttering at high frequencies. As shown, the power falls with increasing frequency following a power-law relationship in all channels, including a visible enhancement near 5.5~mHz (dashed line in Figs.~\ref{fig10} \& \ref{fig11}) in all but the photospheric blue continuum and G-band channels. In general, there is larger power at distances farther from the umbral barycenter across all frequencies. An exception is the power around 5.5~mHz in the Ca{\,}\textsc{ii}{\,}K, H$\alpha$ and IRIS 2796{\,}{\AA} channels, where the power/distance behavior appears to be reversed, i.e., the power near the umbral center is higher than that at the outer radii. This is perhaps similar to the findings of \citet{2012ApJ...746..119R}, where the authors show that oscillations with relatively higher frequencies are more pronounced near the umbral center than at the peripheral regions. \citet{2014RAA....14.1458R} reported similar observations suggesting larger power at high frequencies in a sunspot umbra compared to that at the umbra-penumbra boundary at chromospheric heights. \citet{2012ApJ...746..119R} interpreted this behavior as being due to the differences in inclinations of the magnetic fields across the umbra. However, as one may notice, the Mg{\,}\textsc{i}{\,}b$_{2}$ channel does not show any such enhancement in the power closer to the umbral center, and even more surprisingly, the IRIS 1330{\,}{\AA} transition region channel does not show this either. If it has to do with magnetic field inclinations, one would expect this effect to be at least present, if not more pronounced, in the IRIS 1330{\,}{\AA} channel that captures plasma at more elevated atmospheric heights. This abnormal enhancement near the umbral center at purely chromospheric heights is perhaps caused by a different physical mechanism. \\

\section{Discussion and Conclusions}
Power spectra of a sunspot umbra were generated and studied at multiple atmospheric heights using simultaneous intensity images captured by the ROSA and HARDcam instruments at the DST and by IRIS/SJI. The spectra could be generated up to frequencies as high as a few hundred mHz thanks to the ultra-high cadence observations provided by ROSA/HARDcam. A power-law dependence of the oscillation power is found across the entire frequency range and at all heights, with significant power enhancements found near 5.5{\,}mHz in chromospheric channels. \citet{1997IAUS..181...53G} observed a similar dependence, with an enhancement at 3{\,}mHz found in the power spectrum generated from the line-of-sight velocity oscillations of integrated solar disk light. While the peak at 3{\,}mHz is due to photospheric $p$-modes, the authors describe the power-law dependence at lower frequencies as being due to the random velocity fields introduced by solar convection. However, in the present case, convection is expected to be suppressed in the sunspot umbra, meaning the existence of a similar dependancy at chromospheric heights implies that the fluctuations could be of a different origin. Besides, the same slope at both low and high frequencies suggests that the entire spectrum could actually represent the universal signatures of SMAWs, which cause intensity fluctuations throughout the sunspot umbra. Furthermore, the good resemblance between these power spectra, and those generated from integrated disk light \citep{1997IAUS..181...53G}, confirms the source of sunspot oscillations in the quiet photosphere. The power enhancements near 5.5{\,}mHz highlights the predominance of three-minute oscillations in the umbral atmosphere. 

We also computed phase-difference spectra for different channel pairs, which suggest the oscillations with frequencies up to approximately 3{\,}mHz remain evanescent, while those with higher frequencies propagate universally upwards. The theory of linear wave propagation in a stratified atmosphere, including aspects of radiative cooling \citep{2006ApJ...640.1153C}, agrees well with the phase spectra for the (Mg{\,}\textsc{i}{\,}b$_{2}$, Ca{\,}\textsc{ii}{\,}K) and (Mg{\,}\textsc{i}{\,}b$_{2}$, H$\alpha$) channel pairs. However, the phase spectra for the (Ca{\,}\textsc{ii}{\,}K, H$\alpha$) and (IRIS 2796{\,}{\AA}, 1330{\,}{\AA}) pairs indicate that either the individual channels in each pair form very closely ($<$100{\,}km) to one another within the umbra, or that simple linear wave theory breaks down due to the predominantly nonlinear behavior of waves in these atmospheric regions.

The 3-min ($\approx$5.5{\,}mHz) oscillation amplitudes, as inferred from the normalized power, increase gradually up to the atmospheric height corresponding to the Ca{\,}\textsc{ii}{\,}K observations, before subsequently decreasing beyond. The corresponding energy flux, however, appears to decrease steadily immediately upon leaving the photosphere. Recent observations of upwardly propagating slow magnetoacoustic sausage modes in a magnetic pore also reveal a gradual decrease in energy flux with height \citep{2015ApJ...806..132G}. As can be seen, a slightly steeper reduction in mechanical wave energy is observed above the Ca{\,}\textsc{ii}{\,}K atmospheric height. This is qualitatively similar to the decay in acoustic energy flux calculated from three-dimensional MHD numerical simulations \citep{2011ApJ...735...65F}. The authors describe the decay as being due to a combination of radiative losses and shock dissipation. The lower radiative cooling time in the chromosphere, together with the strong shock features found at Ca{\,}\textsc{ii}{\,}K heights, leads to a steeper decrease in the energy flux at increasing atmospheric heights. Further studies including theoretical and numerical modelling might be useful to ascertain the role of any other damping mechanism(s).

\citet{2011ApJ...735...65F} found an energy flux of the order of 10$^{6}$ erg{\,}cm$^{-2}${\,}s$^{-1}$ at photospheric heights, and concluded that the energy flux available from acoustic oscillations is insufficient to balance the radiative losses in chromospheric umbrae. Recently, using simultaneous observations of a sunspot umbra with Hinode \citep{2007SoPh..243....3K} and IRIS, \citet{2016ApJ...831...24K} estimated the energy fluxes for 6--10{\,}mHz SMAWs at the photospheric and lower transition region levels as 2 $\times$ 10$^{7}$ erg{\,}cm$^{-2}${\,}s$^{-1}$ and 8.3 $\times$ 10$^{4}$ erg{\,}cm$^{-2}${\,}s$^{-1}$, respectively. These results, in contrast to \citet{2011ApJ...735...65F}, demonstrate the potential for SMAWs to contribute significantly to the heating of the umbral chromosphere. The energy flux, in our results, decreases from about 1.3$\pm$0.1 $\times$ 10$^{7}$ erg{\,}cm$^{-2}${\,}s$^{-1}$ at the photosphere to about 40$\pm$3 erg{\,}cm$^{-2}${\,}s$^{-1}$ at the height corresponding to the IRIS 1330{\,}{\AA} channel ($\sim$2000{\,}km). The latter value would be 765$\pm$57 erg{\,}cm$^{-2}${\,}s$^{-1}$ if the alternative height (1500{\,}km), inferred from the phase difference spectra, is assumed for the IRIS 1330{\,}{\AA} channel. In either case, these values indicate significant damping in agreement with \citet{2016ApJ...831...24K}. Of course, it is not trivial to state whether or not the entire missing wave energy directly resulted in the thermalization of the local plasma. For instance, a good fraction of the wave energy could be transferred to other fast/Alfv{\'{e}}n modes through the processes of mode conversion \citep[e.g.,][]{2008SoPh..251..251C, 2013MNRAS.435.2589C}, resulting in dissipationless damping which, in fact, can happen over several scale heights depending on the oscillation frequency. It is not possible to isolate this effect using the current observations. 

The slope of the power spectra (power-law index) is found to increase with height. By comparing the velocity power spectra of propagating kink waves observed in the chromosphere and corona, \citet{2014ApJ...784...29M} demonstrated that there is enhanced damping at high frequencies as the waves propagate towards corona. It is possible that the SMAWs are displaying similar behavior, with stronger damping at higher frequencies, which results in steeper slopes at increased atmospheric heights. Indeed, the oscillation amplitudes above Ca{\,}\textsc{ii}{\,}K height show frequency-dependent damping with shorter damping lengths for higher frequencies (see Fig.~\ref{fig8}). The greater radiative and/or conductive losses for high frequency waves \citep{2002ESASP.505..293C,2012A&A...546A..50K} could perhaps be responsible for their stronger damping in these layers. Additionally, viscosity, ion-neutral collisions, ambipolar diffusion, resonant absorption (via mode conversion) can also produce frequency-dependent damping with more efficiency at high frequencies.

Interestingly, the peak frequency of the 3-min oscillations is found to shift from 5.65 mHz ($\approx$177{\,}s) at Mg{\,}\textsc{i}{\,}b$_{2}$ heights to 6.57 mHz ($\approx$152{\,}s) at Ca{\,}\textsc{ii}{\,}K heights, before remaining at that value for the other chromospheric and transition region channels. This is perhaps a consequence of the variation of acoustic cutoff frequency with height \citep{2016ApJ...819L..23W,2016ApJ...827...37M}. \citet{2016ApJ...819L..23W} performed observations of the quiet solar atmosphere using multiple spectral lines and demonstrated that the acoustic cutoff frequency initially increases with height, before levelling off at greater atmospheric heights. In addition, the presence of harmonic peaks at 6.57{\,}mHz and 13.1{\,}mHz (152{\,}s and 76{\,}s; see Fig.~\ref{fig6}) in the Ca{\,}\textsc{ii}{\,}K and IRIS 2796{\,}{\AA} channels may support the existence of a resonant cavity, possibly below the photosphere \citep{1982SoPh...79...19T}, rather than in the chromosphere \citep{1983SoPh...82..369Z}, since the oscillations are found to be predominantly upwardly propagating at chromospheric heights. This is in contrast to the results of \citet{2015A&A...579A..73M}, where the authors found standing slow modes in a magnetic pore supporting the chromospheric cavity. 

Finally, a comparison of power spectra across the umbral radius reveals an abnormal enhancement in high-frequency ($>$5.5 mHz) power close to the umbral barycenter in chromospheric channels. Although the change in magnetic field inclination angles across the umbra has been shown to produce similar effects \citep{2012ApJ...746..119R, 2015ApJ...800..129M}, the restriction of this behavior in our current observations to chromospheric channels (i.e., excluding the transition region observations) is puzzling and demands further exploration.

\acknowledgements 
The authors thank the referee for useful comments. SKP is grateful to the UK Science and Technology Facilities Council (STFC) for funding support that allowed this project to be undertaken. DBJ also wishes to thank the UK STFC for the award of an Ernest Rutherford Fellowship, in addition to a dedicated research grant. DBJ is also grateful to Invest NI and Randox Laboratories Ltd. for the award of a Research \& Development Grant (059RDEN-1). TVD was supported by an Odysseus grant of the FWO Vlaanderen, the IAP P7/08 CHARM (Belspo) and the GOA-2015-014 (KU~Leuven). This work was based on discussions at the ISSI and ISSI-Beijing. This project has received funding from the European Research Council (ERC) under the European Union's Horizon 2020 research and innovation programme (grant agreement No 724326). RM is grateful for the support of a Leverhulme Trust Early Career Fellowship. VF would like to thank the STFC and Royal Society-Newton Mobility Grant NI160149, for their financial support. RE is grateful to STFC (UK) and acknowledges The Royal Society (UK) for the support received. IRIS is a NASA small explorer mission developed and operated by LMSAL with mission operations executed at NASA Ames Research center and major contributions to downlink communications funded by ESA and the Norwegian Space Centre. 


\end{document}